\documentclass[a4paper]{jpconf}
\usepackage{graphicx}
\usepackage{bm}
\usepackage{multicol}
\usepackage{multirow}

\begin{document}
\title{YUI and HANA: Control and Visualization Programs for HRC in J-PARC}

\author{Daichi~Kawana$^1$, Minoru~Soda$^{1,}$\footnote[3]{Present address: Center for Emergent Matter Science, RIKEN, JAPAN.}, Masahiro~Yoshida$^1$, Yoichi~Ikeda$^{1,}$\footnote[4]{Present address: Institute for Material Research, Tohoku University, JAPAN.}, Toshio~Asami$^1$, Ryosuke~Sugiura$^1$, Hideki~Yoshizawa$^1$, Takatsugu~Masuda$^1$, Takafumi~Hawai$^2$, Soshi~Ibuka$^2$, Tetsuya~Yokoo$^2$, and Shinichi~Itoh$^2$}

\address{$^1$ Neutron Science Laboratory, Institute for Solid State Physics, the University of Tokyo, JAPAN}
\address{$^2$ Neutron Science Division, Institute of Materials Structure Science, KEK, JAPAN}

\ead{kawana@issp.u-tokyo.ac.jp}

\begin{abstract}
We developed control and visualization programs, YUI and HANA, for High-Resolution Chopper spectrometer (HRC) installed at BL12 in MLF, J-PARC.
YUI is a comprehensive program to control DAQ-middleware, the accessories, and sample environment devices.
HANA is a program for the data transformation and visualization of inelastic neutron scattering spectra.
In this paper, we describe the basic system structures and unique functions of these programs from the viewpoint of users.
\end{abstract}

\section{Introduction}

High Resolution Chopper spectrometer (HRC) is an inelastic neutron chopper spectrometer to investigate elementary excitations of spin and orbital fluctuations in strongly correlated electron systems with high energy resolution.
Recently it is used also for the study of hydrogen atoms in a solid. 
HRC is installed at BL12, in Material and Life-science experimental Facility (MLF), J-PARC~\cite{Itoh_11, Itoh_13}.
Basic accessories including vacuum chamber, T0 chopper, Fermi chopper and soller collimator, and minimal sample-environmental devices 
had been prepared until 2010, and then we have constructed computing system and developed software environment to control these devices and to handle neutron signals obtained from detectors.
Comprehensive-control program YUI and visualization program HANA have central roles in the software environment.
In this paper, we firstly explain the configuration of the computing-system which underlies the two programs.
Next, we explain basic constructs and unique and characteristic functions in YUI and HANA.


\section{System Configuration}

\begin{figure}[p]
\centering
\includegraphics[width=45pc, angle=-90]{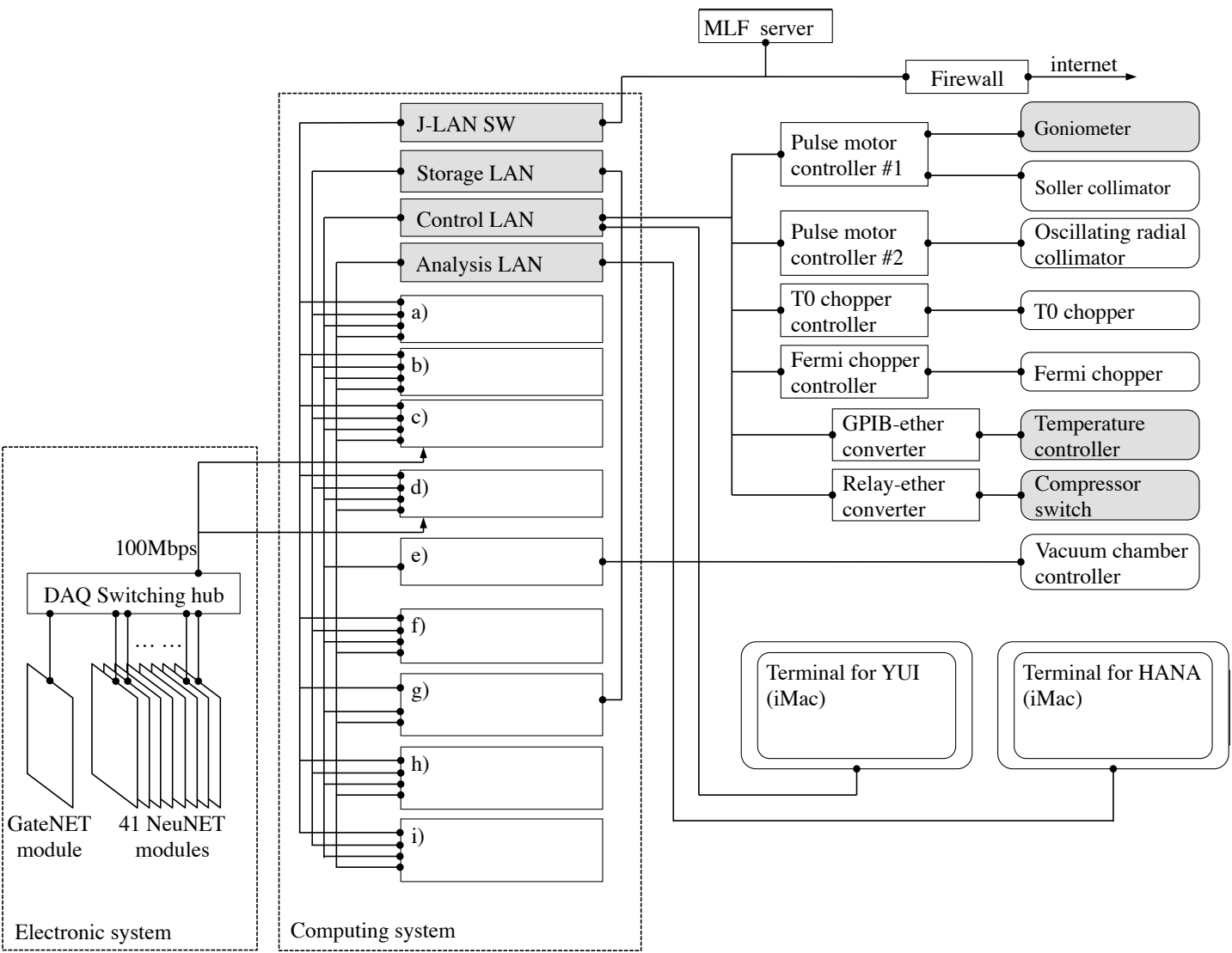}
\caption{\label{fig:machines}Schematic diagram of computing system in HRC. 
The indices a)-i) denote PC machines: a) versatile server, b) DAQ operator, c)-d) 2 DAQ CPUs e) Gateway for FL-net, f) server for analysis, and g)-i) 3 denote NAS's. Their functions and roles are described in the text. Open and hatched round-squares denote accessories and sample-environment devices, respectively.}
\end{figure}

The schematic view of the computing system in HRC is shown in figure~\ref{fig:machines}.
We use a Scientific Linux-based computing system to run the Data AcQuisition (DAQ) 
middleware~\cite{SSatoh_09}, which is used in other beamlines of MLF as well. 
It operates the dedicated programs necessary to perform user experiments at HRC.
The roles of the computing machines in HRC are summarized in table~\ref{tab:machines}.
They are linked via four types of intranet networks: storage LAN, control LAN, analysis LAN, and J-LAN.

The storage LAN directly handles the event data.
An electric signal ignited by a neutron detection at a position sensitive detector (PSD) is transformed to the event data of an 8-bytes binary data-set on the corresponding NeuNET module~\cite{Yano_11}.
Two DAQ CPU machines work as the gateway.
The number of the machines is determined by taking into account the total number of PSDs (324 PSDs as in Mar.~2017).
The obtained event data are stored to a network-attached storage (NAS) machine.
We have also set up two more storage machines, which work as archives for all the event data since the start of HRC operation in 2010.
The control LAN intervenes between the computing system and all the controllable devices. The program YUI runs on a DAQ-operation machine, and it operates DAQ-middleware and controls these devices.
At the same time, a versatile server machine performs the aggregation and logging of status from the devices.
The role of the ``device server'' of the server will be explained in the following section. 
The analysis LAN processes the data visualization from the raw event data preserved on the NAS.
The program HANA has been developed as the user interface for data analysis.
All the machines connect to J-LAN, the standard intranet in MLF, necessary to get the proton number, as mentioned later.

The computing system in HRC is based on the distributed computing, and we adopted a client-server model.
Thus, YUI and HANA are available as the client terminal in two iMac PCs at the user's room in the HRC cabin. 
Users do not need to operate the above Linux-machines directly.

\begin{table}[h]
\caption{\label{tab:machines}Role of each machine installed in HRC. The same indices are used in figure~\ref{fig:machines}}
\begin{center}
\begin{tabular}{llp{100mm}}
\br
Index & Machine & Role and function \\
\mr
a) & Versatile server & Installation of device servers and handling of logging data \\
b) & DAQ operator & Operation of DAQ, CPU of YUI \\
c) & DAQ CPU\#1 & Collection of event data from PSDs \\
d) & DAQ CPU\#2 & {\it Ditto} \\
e) & Gateway for FL-net & Gateway to communicate with the vacuum-chamber controller \\
f) & Server for analysis & CPU of HANA \\
g) & NAS\#1 & Storage for event-data of recent several years and for analyzed data by users \\
h) & NAS\#2 & Backup of g) and storage for all the event data \\
i) & NAS\#3 & Backup of h) \\
j) & Terminal for YUI & Terminal for YUI \\
k) & Terminal for HANA & Terminal for HANA \\
\br
\end{tabular}
\end{center}
\end{table}

\section{YUI}

\begin{figure}[h]
\centering
\includegraphics[width=36pc]{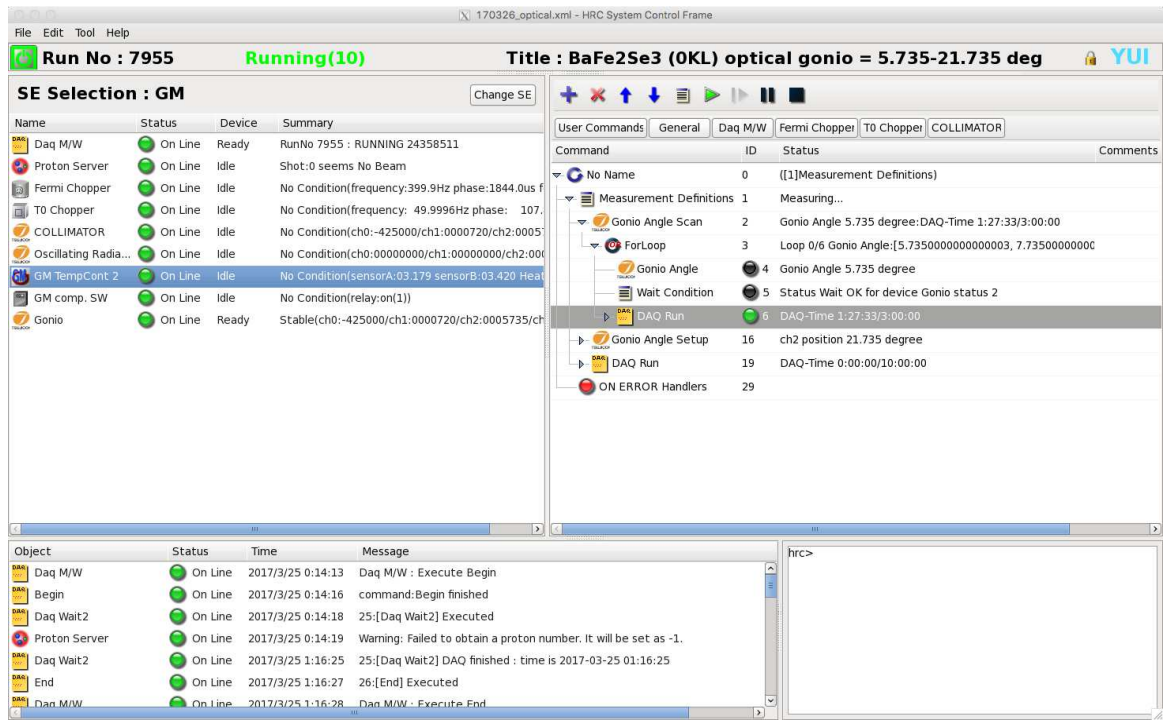}
\caption{\label{fig:YUImain}The main window of YUI. It mainly consists of three display-areas: list of activated devices (top-left), sequence edited by users and its progress status (top-right), and messages communicated with devices, called system log (bottom).}
\end{figure}

We have named a comprehensive control program as YUI. The capital Y is from the first character of 
Japanese word {\it Yasashii} (friendly or easy) or {\it Yoku-dekita} (well-constructed). 
YUI is the abbreviation of Yasashii User Interface or Yoku-dekita User Interface.
YUI is an integrated program that allows users to perform the measurement by sequential operation of DAQ-middleware to control the devices, and to log their status. The main window of YUI is shown in figure~\ref{fig:YUImain}.
The feature of YUI is that all the related devices in HRC are controlled from the single program YUI 
to provide users with a comfortable experimental environment, 
although their communication protocols are different among them.
As a key-solution, ``device server,'' a virtual machine which intervenes between YUI and the corresponding device, is developed and installed in the versatile server machine.

The devices that can be controlled from YUI are summarized in table~\ref{tab:devices}.
All the devices are categorized into two types; accessory devices that are commonly used for all the 
experiments, and sample-environment devices such as the GM-type refrigerator and 
$^3$He circulation-type refrigerator~\cite{Itoh_16}. 
In the latter one device is selected for each experiment.
At HRC, each sample-environment device has individual temperature controller and goniometer.
When starting experiment, users select one sample-environment as a mode of YUI, 
so that the appropriate devices are automatically enabled.
As the result, the available devices on YUI can be changed easily and smoothly
when the sample-environment device is changed.

\begin{figure}[h]
\centering
\includegraphics[width=24pc]{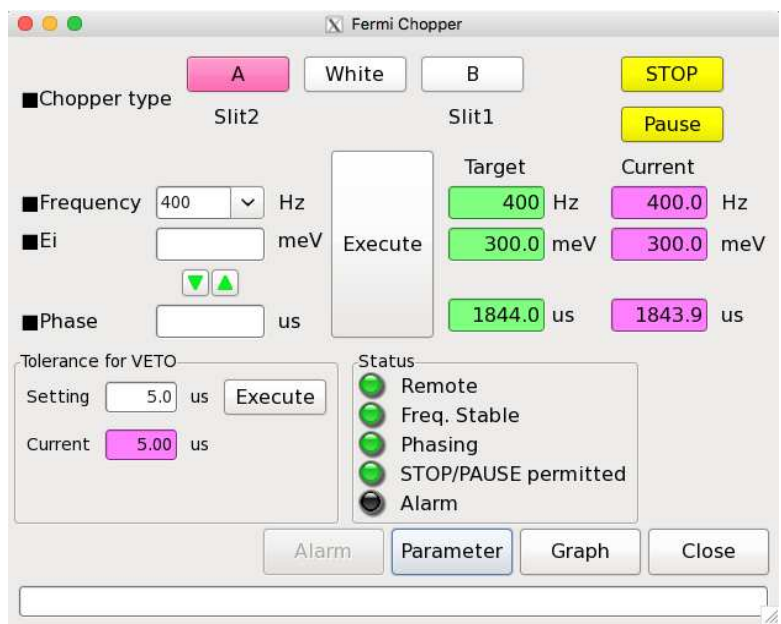}
\caption{\label{fig:Fermi}The GUI for the control of the Fermi chopper.}
\end{figure}

At YUI, user can control devices in either of two ways. One is the manual control by operating the devices directly.
For each device, we have prepared the specialized control graphical user interface (GUI). 
As an example, the control GUI of Fermi chopper is shown in figure~\ref{fig:Fermi}.
This GUI allows users to execute all commands necessary for experiments such as selection 
of one chopper from two ones mounted on the chopper stage and setting of 
values of frequency and incident energy.

The other way is automatic control by a macro of sequential commands.
The macro can be edited by users, but built-in macros called 
``user-commands'' are prepared so that even a first-time user can operate YUI easily.
Users can perform their experiment by just arranging these user-commands in a sequence file. 
The edited sequence file can be saved as xml format for the convenience. User-command 
consists of a set of basic commands which are prepared for each device. In usual experiments, 
the order of basic commands for the user-command never changes. For example, the ``DAQ Run'' 
as mentioned above is one of the typical user-commands. While executed DAQ Run, the 
basic commands derived from DAQ-middleware such as configure, begin, DAQ wait, end unconfigure, 
and increment run number are executed in sequence. Furthermore, as an application of user commands, 
``Gonio-angle-scan'' is a convenient user-command that consists of DAQ Run and a series of basic commands related to goniometer control.

On YUI, user mode and expert mode are implemented.
The user mode is good for most of experiments of general user program in case that experimental conditions do not have to be changed.
Users should note that they cannot access the detail setup of devices.
In the expert mode, flexibility of making macro is very high. 
Following operations are performed in the mode; 
setting of the phase offset of Fermi chopper in the initial operations after long shutdown, 
special operation which is not assumed in user program of Fermi chopper to confirm the stability on its phase setting, 
unusual activation of the sample-environment devices due to a trouble of the device, 
and sequence test performed when YUI is updated.
In the expert mode, any order of the basic commands is possible, and improper order may cause unexpected behavior.
Careful handling is required. 

The current values of the devices are continuously monitored, and they are written in the log files every 5~seconds so that users know whether the experiment is working or not.
Additionally, Log-visualizer on a browser page helps users to check the time variation of these values.

After a measurement, a report file is created.
From the log data, the set and monitored values of all the devices at the time of beginning and completion of the measurement are recorded.
In addition, effective measurement time and proton number which is obtained from the server in MLF intranet~\cite{PNserver} are also recorded.
The proton number is used for normalization of the measured intensity.

\begin{table}
\caption{\label{tab:devices}Installed devices to be handled by YUI.}
\begin{center}
\begin{tabular}{p{30mm}p{30mm}p{30mm}p{55mm}}
\br
Device Category&Device&Communication port and interface& Logging parameter(s)\\
\mr
\multicolumn{4}{l}{Accessory (common use for all the experiment)}\\
\mr
& Vacuum Chamber&FL-net gateway&(Monitoring only)\\
&&&- Pressure\\
&&&- Valve status (Open/Close)\\
&T0 chopper&MELSEC&(Monitoring only)\\
&&&- Frequency\\
&&&- Phase (delay)\\
&&&- Temperatures\\
&DAQ middleware&(HRC system)&Begin/end time of measurement\\
&MLF-server&J-LAN Intranet&Proton current, converted into the amount of monitored incident neutrons.\\
&Fermi chopper&MELSEC&- Selection (A, B, or none)\\
&&&- Frequency\\
&&&- Phase\\
&&&- Tolerance\\
&Soller collimator&\multirow{4}{30mm}{Pulse motor controller (Tsuji-con PM-16C)}&- $z$-axis (types of collimator)\\
&&&- $\omega$-axis (for alignment)\\
&&&\\
&&&\\
&Oscillating radial collimator&Pulse motor controller&- $\omega$-axis (Oscillation on/off or non-use)\\
\mr
\multicolumn{4}{l}{Sample-environment device}\\
\mr
GM refrigerator&Goniometer&Pulse motor controller&- $\omega$-axis (sample orientation in plane)\\
&Temperature controller (Lakeshore 335)&GP-IB converter to TCP/IP&- Temperature\\
&Power switch for compressor&Relay converter to TCP/IP&Power switch for Compressor\\
\multirow{2}{30mm}{$^3$He circulation type refrigerator}&Goniometer&\multirow{2}{30mm}{Pulse motor
controller}&- $\omega$-axis (sample orientation in plane)\\
&&&\\
&Temperature controller (Lakeshore 335)&GP-IB converter to TCP/IP&- Temperature\\
\multirow{2}{30mm}{Superconducting magnet}&\multirow{2}{30mm}{Superconducting magnet}&\multirow{2}{30mm}{(Miscellaneous) IROHA2 based}&- Temperature\\
&&&- Magnetic field\\
&&&- Angle of sample stick\\
\br
\end{tabular}
\end{center}
\end{table}

\section{HANA}

HANA (HRC ANAlyzer) is a program for the initial data handling and the data visualization.
The basic flow is the creation of intensity-histogram data from the event data, sequent reduction of 
the histogram data, and the visualization 
of the neutron cross section as a two-dimensional intensity map.

The event data obtained through DAQ-middleware includes a parameter set 
$\{d; Q_1, Q_2$; $t\}$, where $d$ is PSD number, $Q_1$ and $Q_2$ are 
electric pulse-heights at the upper and lower ends of PSD, respectively, 
and $t$ is time of flight (ToF) of neutron~\cite{SSatoh_09}.
Here the ToF is the time between the generation of a neutron at 
the decoupled target and the reach of the neutron on PSD.
The values $Q_1$ and $Q_2$ give the coordinate of the neutron on a 
PSD, $p$, which is represented as an integer in the range of $0 \le p \le 1023$.
Here 0 and 1023 correspond to the bottom and top of the PSD, respectively.
The relation between $p$, $Q_1$, and $Q_2$ is as follows; 
$p=1024\left[a Q_2/(Q_1+bQ_2)-c\right]$, where $a, b$, and $c$ are 
position-correction parameters.
Thus, the array of PSDs is regarded as a two-dimensional screen having $\{d,p\}$ pixels.
This means that one detected neutron has three parameters of $\{d,p,t\}$.
Event data obtained in one measurement are transformed into histogram data in this way.

The next process is the transformation from $\{d,p,t\}$ to $\{d,p,\omega\}$, 
where $\omega$ represents the energy transfer of the neutron.
In prior to the estimate of $\omega$, the incident energy $E_i$ is 
obtained by using ToF of neutrons, $L_1$, and $L_2$. 
Here $L_1$ is the distance between the target and the center of sample ($L_1$=15~{\rm m} for HRC), 
and $L_2$ is the distance between the center of sample and the position on PSD.
Then, using the value of $E_i$ and the parameters of the event data $\{d,p,t\}$, $\omega$ is obtained 
using the following relation; 
\begin{equation}
	\omega= E_i-\frac{1}{2}m\left[ \frac{L_2}{t- L_1/\sqrt{2E_i/m}}\right]^2,
\end{equation}
where $m$ is the mass of neutron.
Thus, the event data of $\{d,p,t\}$ is transformed into the form of $\{d,p,\omega\}$, and the histogram data $I(d, p, \omega)$ is created.


The method of the transformation from $\{d,p,t\}$ to the $\{\omega , \bm{Q}\}$, where ${\bm Q}$ is momentum transfer of neutron, is 
different between the case of single-crystal and that of powder samples. For single-crystal sample, users must give lattice parameters $a, b, c; \alpha, \beta, \gamma$. In addition, a vertical-axis rotation angle $\Psi$, corresponding to the set angle of the goniometer, a slope angle from the vertical axis $\delta$, and its direction $\tau$ are also given with consideration for the arbitrary orientation of the single crystal. Using these parameters, $\bm{Q}$ is represented by the coordinate of the reciprocal lattice $Q_a,~ Q_b,~ Q_c,$ as shown in figure 4. In the intensity map of HANA, two parameters are selected from the 
$Q_a, Q_b, Q_c$, and $\omega$ for the vertical and horizontal axes. The other two parameters are integrated. In case of powder sample, the relation between the scattering angle $2\theta$ of a neutron and $|\bm{Q}|=Q$ is calculated by
\begin{equation}
  Q^2= G \left[ 2Ei-\omega -2\sqrt{Ei(Ei-\omega)}\cos{2\theta} \right],
\end{equation}
where $G (=2m/\hbar^2)$ is constant. 

The histogram data $N(d,p, \omega)$ can be transformed to a data format of a visualization program MSlice~\cite{Mslice} 
for the convenience of users. Several data with different measurement times can be combined by normalization using proton number recorded by YUI. 

We have implemented some unique functions in HANA. One of them is the alignment support function, which is indispensable in the beginning of single crystalline experiments.
The sequence is as follows.
First, Laue profile is measured using white beam, and the intensity map $I(d,p)$ is displayed.
The precise position of $\{d,p\}$ for a focused Laue spot is determined by fitting the intensity using Gaussian function.
Then, the intensity at the position is resolved into time, and it is plotted as a function of ToF.
A set of Bragg peaks including higher-order reflections are shown. By using the distance from sample to the position of $\{d,p\}$, the ToFs are converted to scalers of reciprocal lattice vectors ($d^*$s).
Users, thus, can index the Bragg reflections, and they obtain the information of the relation between the coordinate of the crystal and that of the spectrometer. 

In general, the detection efficiency in the $\{d,p\}$ map has a distribution because 
of the character of PSD and the variation of the solid angle along the perpendicular direction. To correct these effects, we use an intensity map of a standard 
Vanadium sample $V(d,p)$. It is measured using white beam at a commissioning. The corrected histogram data is obtained as follows; 
$I_1(d,p,\omega)= V_0 I_0(d,p,\omega)/V(d,p)$. 
Here the $I_0(d,p,\omega)$ is the data of the user and $V_0$ is a constant so that the magnitudes of 
$I_0$ and $I_1$ are close. 
We have also developed the distribution versions of HANA, for Win and Mac. They are distributed for free. 
Users are able to perform data reduction process using event data in their own campus. 
\begin{figure}[h]
\centering
\includegraphics[width=24pc]{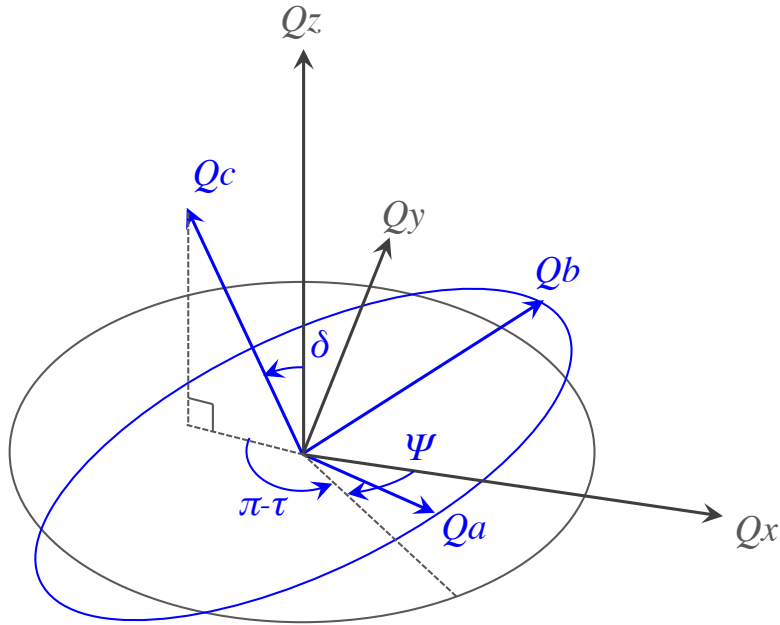}
\caption{\label{fig:config}Definitions of $Q_a, Q_b$, and $Q_c$.  The coordinate $\{Q_x, Q_y, Q_z\}$ reprsents the detector coordinate, where $Q_x$ is parallel to the incident beam, $Q_y$ is perpendicular to the beam within a horizontal plane, and $Q_z$ is vertical to the plane.}
\end{figure}


\section{Summary}

We have developed the control program YUI and the data-analysis program HANA 
for HRC spectrometer. 
They have been improved day-by-day for convenience of users and instrument staff members. 
The hardware of HRC is always updated; for example a new soller collimator has been installed for more neutron flux, 
and a new device for sample environment will be installed in future. 
YUI also evolves with the progress of the hardware. 
On the other hand in HANA, we are planning to implement a functions of 
the analysis for four-dimensional data obtained from the continuously rotation scan of a single crystal.


\ack{The development of the computing system and software environment for HRC are approved and progressed by the Neutron Scattering Program Advisory Committee of IMSS, KEK (Proposal No. 2010S01-2016S01). We would like to appreciate Bee Beans Technologies Co.,Ltd. (Tsukuba, JAPAN) for gratefully supporting our works as partnership.}


\end{document}